\documentclass[12pt,a4paper,fleqn]{article}
\usepackage{amsmath}

\usepackage{graphicx}
\usepackage[]{amsmath}
\usepackage{textcomp}

\usepackage{parskip}
\usepackage{setspace}
\usepackage{verbatim}
\usepackage{subfig}
\usepackage{cite}
\usepackage{url}
\usepackage{longtable}
\usepackage{multicol}
\usepackage{multirow}

\usepackage{amssymb,amsfonts,textcomp}
\usepackage[margin=1.0in]{geometry}

\DeclareGraphicsExtensions{.eps}

\newcommand{\be}{\begin{equation}}
\newcommand{\ee}{\end{equation}}

\title{A method for importance sampling through Markov chain Monte Carlo with post sampling variational estimate }
\author{A. John Arul \\
\footnotesize Indira Gandhi Centre for Atomic Research,  Kalpakkam, India\\
\footnotesize \texttt{arul@igcar.gov.in}
\and
Kannan Iyer\\
\footnotesize Indian Institute of Technology, Bombay, Mumbai, India\\
}
\date{Dec. 2013}

\begin{document}
\maketitle

\abstract{We propose a method to efficiently integrate truncated probability densities. The method 
uses Markov chain Monte Carlo method to sample from a probability density matching the function being 
integrated. The required normalisation 
or equivalently the result is obtained by constructing a function
with known integral, through non-parametric  kernel density estimation and variational procedure. 
The method is demonstrated with numerical case studies.
Possible enhancements to the method and limitations are discussed.}


\section{Introduction}

In importance sampling Monte-Carlo methods, the variance  reduction strategy  is implemented by
choosing a sampling density which closely mimics the function to be integrated.
The new sampling density also weights the original function to correct for
this introduced sampling bias~\cite{Liu}. Given that an optimal sampling 
density might exist, leads naturally to the development of algorithms, which 
optimize the sampling density dynamically from the past samples. 
These are known as adaptive importance sampling and a number of
schemes have been proposed in the literature for enhancing the efficiency 
of importance sampling methods.   These methods, variational or otherwise
operate on the sampling function to optimize it, with information derived from
the samples sequentially.  

In this study we propose a method by which, the samples are generated 
exactly as per the function to be integrated (function of interest ) 
through Markov Chain Monte Carlo (MCMC) method.
Here 'exactly' means the sampling density differs from the 
function to be integrated by a constant factor. This is the condition for 
minimum variance ~\cite{Liu}. In the next step of the method, a weighting function of 
known integral is constructed from the samples generated in the previous step. 
In the third step, the integral is evaluated as an expectation of the
ratio of the weighting function and function of interest or by a  
variational principle which minimizes the distance between the function of interest 
and weighting function.  The steps two and three could be combined for the 
minimization of variance. 

Though Monte-Carlo methods are useful for integration of high dimensional 
integrals, we demonstrate the feasibility of the proposed method
by application to simple one dimensional integrals, that are typical 
in exceedence probability estimates.   The report is organised as follows.   
Section 2, reviews the importance sampling Monte-Carlo principle
and adaptive importance sampling methods.  
In section 3, the proposed method is presented. The subsections discuss the 
techniques such as  MCMC and Kernel Density Estimation (KDE) essential in the proposed
method. The  use of Hybrid Monte-Carlo (HMC) for improving the 
performance characteristics is also presented in the subsections, as the function to be integrated
has non-smooth regions.  Results of numerical experiments with the proposed method is 
presented in section 4, where 
the performance is compared with direct or analogue MCS and other 
known importance sampling algorithms like simple importance 
sampling Monte-Carlo Simulation (ISMCS) and 
concluded in section 5.

\section{Importance sampling}

A number of schemes have been proposed in the literature to effect variance reduction. The following is a list of techniques ~\cite{Liu, Whitlock} to name a few.
\begin{itemize}
\item
Stratified sampling: Trial density is constructed as piecewise constant function.
\item
Use of negatively correlated sampling or antithetical variables
\item
Divide and conquer approaches: Divide the region of integration into different regions 
where different sampling functions can be used.
\item
Subset simulation: Divide and conquer strategy combined with Bayesian method, where the region of interest is represented by a sequence of subsets, whose intersection progressively results in a closer approximation to the actual domain \cite{Au2001}.
\item
Variational and Bayesian adaptive importance sampling: The weighting function is adjusted by a variational method or Bayesian inference, to minimize variance \cite{Roche}. 
\item
Quasi-Monte Carlo: Uses well distributed deterministic sequences instead of pseudo random numbers 
\cite{Whitlock}.  
\end{itemize}

Importance sampling is another general and well known class of variance reduction technique. 
See \cite{Whitlock, Liu} for a good introduction and \cite{Laura} for 
a recent discussion on the limitations. The general approach to importance 
sampling is as follows.
The integral of a function $f(x)$,  in the domain $[a,b]$,  
\begin{equation}
I=\int_{a}^{b}{f(x)  dx}
\label{eq:int}
\end{equation}
can be evaluated as an expectation of f(x), with respect to a $pdf$, $\rho(x)$ as,
\begin{eqnarray}
\label{eq:isw}
I= \int_{a}^{b} \frac{f(x)}{\rho(x)} \rho(x) dx 	\\
= <\frac{f(x)}{\rho(x)}>_{\rho}	
\end{eqnarray}
where $<f(x)>$, is the statistical expectation value 
\be
=\frac{1}{N} \sum_{i=1}^N \frac{f(x_i)}{\rho(x_i)}
\ee
There is freedom in choosing $\rho$, to generate the
samples  $x_i$.  The well known variance reduction algorithms optimize $\rho$, to yield minimum
variance. 

The standard result in variance reduction can be obtained as in ~\cite{Whitlock}, from the expression for the variance of I, 
\begin{equation}
var(I)= \int \left[ \frac{ f(x)}{ \rho(x)} \right ]^2 \rho(x)dx \,-\, I^2 
\end{equation}
From this equation one can infer the condition for minimum variance as,
\begin{equation}
f(x)  \lambda =  \rho(x)
\label{eq:cond-mv}
\end{equation}

where  the constant, $\lambda = I^{-1}$.
Therefore, the condition for minimum variance \cite{Whitlock}, is that the sampling density $\rho$ is a constant multiple of the function $f(x)$ to be integrated.
The variance reduction principle implies that, a  Monte Carlo estimate can be made more accurate for a given number of samples (or equivalently variance reduced) by  sampling more frequently from the region where the
function is maximum. In a sense this implies that more the information
one has about the function to be integrated, smaller will be the 
variance. 

In the next section a variant of adaptive importance sampling is proposed. 
Here the function to be integrated is cast as the weighting function in the
integral to evaluate a function of known integral . This function of known 
integral is optimized with respect to the samples, to minimize variance. 
The new normalised function introduced takes the role of the weighting function, but the sampling
is directed by the original $pdf$ whose integral is to be evaluated.  

\section{Post sampling variational  MCS} 
\label{sec:psam}
 
In the field of reliability and risk analysis, it is often required to compute integrals of the form, 
\begin{equation}
I = \int {f(x) C(x)} dx   
\label{eq:rel-eq}
\end{equation}
Here $C(x)$ is a criteria function ($x$ could be a vector, $x \in R^n$), which distinguishes 
the success and failure domains ($\Omega_f$). The variables $x$ are the parameters in the
dynamical system which are considered to be uncertain characterised by the $pdf$,  $f(x)$.  
The form of $C(x)$ is,

\begin{equation}
C(x) = \left\{ \begin{array}{ll}
                1 & {when \, x \in \Omega_f}  \\ 
                0 & {otherwise}  
                \end{array}
                \right.
\end{equation}
 $C(x)$ is usually implemented by a computer code. We can 
cast this integral as in \ref{eq:isw}  with a weighting $pdf$ , $g(x)$.
\begin{equation}
I= \int \left [ \frac{ f(x) C(x)}{ g(x)} \right ] g(x)dx  
\end{equation}

From equation \ref{eq:cond-mv} the condition for minimum variance implies,
\begin{equation}
g(x) =  \lambda f(x) C(x)
\label{eq:mv-cond}
\end{equation}
with $\lambda = I^{-1}$. Therefore, in this equation we can minimize the variance by choosing  $g(x) \sim C(x)f(x)$. i.e.,
$$g(x) =  \lambda C(x) f(x) = \lambda g'(x)$$
$\lambda$ is a normalizing constant of function $g'(x)$, and $\frac{1}{\lambda}$gives the required probability integral of $g'(x)$. In problems where $g'(x)$ is known without the  normalising constant, as in reliability problems ($\lambda$ is the required solution) the following
approach is taken. Let h(x) be a normalised 
function such that $I= \int h(x)d(x) = 1$. We can re-write this integral as follows, 
\be
I = \int\frac{h(x)}{\lambda g'(x)}g(x)dx = 1
\label{eq:10}
\ee
with  $g'(x)$ denoting the reliability function to be integrated as,
$g'(x) = C(x)f(x)$. Rearranging \ref{eq:10} for $\lambda$,
\be
\lambda = \int\frac{h(x)}{g'(x)}g(x)dx
\ee
with the condition, $h(x) = 0$ whenever $g'(x)=0$. Therefore, $\lambda$
can be evaluated as the expectation,
\be
\lambda = \left<\frac{h(x)}{g'(x)} \right>_g
\ee
\be
\lambda=\frac{1}{N}\Sigma_i\frac{h(x_i)}{g'(x_i)}
\label{eq:20}
\ee
Expressing the integral in this form, with the function to be chosen as $h(x)$ instead of 
the probability function $g(x)$, provides flexibility to optimize $h(x)$, after the samples 
have been chosen, and hence is referred to as post sampling variance reduction 
in this report. 

The algorithm works as follows.
Given a probability integration problem, with f(x) denoting the 
joint density in the input parameter space and C(x), the indicator
function computed using a numerical code or analytical function,

\begin{enumerate}
\item generate samples from g'(x) using MCMC algorithm ~\cite{Metropolis}.
\item construct a $h(x)$ which minimizes the variance of eq. $\ref{eq:20}$.
\item perform the sum to evaluate $\lambda$, which is the required result.
\end{enumerate}

\subsection{Markov Chain Monte Carlo}
 
The  MCMC is a general method of obtaining samples from a given distribution.
This  method is the choice when the normalising constant of the $pdf$ of interest is unknown.
A Markov chain is a stochastic process where the next state is determined only by the
current state and transition probability. Given a transition rule, there is an invariant density 
corresponding to the Markov process defined by the
transition rule. Metropolis \cite{Metropolis} gave a prescription for obtaining the transition 
rule corresponding to a given $pdf$ (invariant density). 
   
The metropolis algorithm is as follows ~\cite{Gilks1996},
Select an initial parameter vector $x_0$. Draw trial step from a symmetric 
pdf, i.e., $p(\Delta x) = p(-\Delta x)$ \\
Iterate as follows: at iteration number k,
\begin{enumerate}
\item create new trial position $x' = x_k + \Delta x$,
where $\Delta x$ is randomly chosen from $p(x)$.
\item calculate the ratio $r = \frac{\pi(x' )}{\pi(x_k)}$
\item calculate acceptance probability $\alpha = min\{1,r\}$
\item set  $x_{k+1} = x'$ with probability $\alpha$, set $x_{k+1} = x_{k}$ 
with probability $1- \alpha$.      
\end{enumerate}

Here $\pi(x)$ is the density we want to sample from and $p(x)$ is a proposal 
density, which is often taken as Gaussian or similar to that of the target density.  
Hastings generalised this procedure, by introducing non-symmetric proposals. The modified acceptance criteria, known as Metropolis-Hastings \cite{Liu}, is
\be
\alpha= min \left \lbrace 1,\frac{\pi(x') T(x',x) }{\pi(x) T(x,x')} \right\rbrace
\ee
Here, T(x,x') is the transition probability from $x$ to $x'$. For proposals with
symmetric transition probability, the above formula reduces to Metropolis acceptance criteria.

The above algorithm is referred to as symmetric random walk metropolis algorithm \cite{Andrieu2008}
and is easy to implement. The MCMC method may exhibit slow convergence to target distribution ~\cite{Neal}. 
This has prompted various approaches to improve convergence, such as perfect sampling,
burn-in, coupling from the past, 
Hybrid Monte Carlo. These are some of the techniques for accelerating mixing and convergence
of plain MCMC.  Adapting the parametrized proposal density  (transition probability) is also a 
line of methods known as adaptive MCMC  for improving the efficiency as discussed in the next subsection. 

\subsection{Adaptive Importance sampling} 

In MCMC the proposal function can be optimized instead of manual tuning, for 
efficient generation of samples. The methods developed to optimize the parameters 
of the proposal distribution from the samples are referred to as adaptive MCMC
in the literature. See for instance \cite{Andrieu2008} for a discussion on optimizing
parametrized transition function of MCMC.  In the context of importance sampling, independent of the 
sampling strategy,  the  weighting function can also be optimized to yield
minimum variance as the sampling progresses.  See for instance 
\cite{Roche} for the use of variational sampling to optimize approximations to the sampling density
in the area of parameter inference.
In the next  section a technique for enhancing the mixing characteristics of Markov chain is discussed
for use in the proposed method.

\subsection{Hamiltonian Markov Chain Monte-Carlo}

To improve the mixing time or burn in time of the MCMC sequence, Hamiltonian MCMC (HMC) is 
used in this study. This section presents the basic aspects
of HMC ~\cite{Neal}. This method is also known as Hybrid Monte-Carlo. 
The first step in the method is to define
an energy function (Hamiltonian) involving the variables  which needs to be sampled from  given distributions. In addition for each of the variables to be sampled an auxiliary variable is defined, which typically  have independent  Gaussian distribution. The Hamiltonian is defined as,

$$
H=U(q)+K(p)
$$

The kinetic energy $K(p)$, in terms of momentum variable is  $= p^2/(2m)$.
The potential energy $U(q)$ is taken as the negative logarithm of the probability
density for the variables  $q$.

Then, an evolution of the variables in the phase space 
is defined in terms of the Hamiltonian dynamics. Hamiltonian dynamics involving position like  
variables $q$ and momentum like variables $p$ are,

\begin{eqnarray}
\frac{dq}{dt}=\frac{\partial{H}}{\partial{p}} \\
\frac{dp}{dt}=-\frac{\partial{H}}{\partial{q}}
\end {eqnarray}

where, $H$ is known as the Hamiltonian (Energy function). 
The numerical solution is obtained from the Leapfrog's method \cite{Neal}, where the solutions
are obtained on staggered grid, as follows,

\begin{eqnarray}
p(t+\epsilon/2) = p(t)-\frac{\epsilon}{2} \frac{\partial U}{\partial q} \vert_{t}  \\
q(t+\epsilon)=q(t)+\epsilon  \frac{\partial K}{\partial p}	\vert_{t+\frac{\epsilon}{2}} 	   \\
p(t+\epsilon)=p(t+\frac{\epsilon }{2})-\frac{\epsilon}{2}\frac{\partial U}{\partial q} \vert_{t+\epsilon} \\
\end{eqnarray}
	
For the probability integral problem, the Hamiltonian energy function $H=K(p)+U(q)$ is defined as follows,
\be
 H(q,p)=\frac{p^2}{2 m}-log(g(q))	
\label{eq:leap-frog}
\ee

The acceptance probability $\alpha$  for the new variables $(p',q')$ is,
$$
\alpha = min \left \lbrace 1,\frac{exp(-H(p',q')) }{exp(-H(p,q))} \right\rbrace
$$

The HMC algorithm has the following  steps.
\begin{enumerate}
\item make a new proposal as per the first step of MCMC, say $q$.
\item perform $n=\ell/\epsilon$ steps of Hamiltonian evolution with pseudo momentum $p$.
\item calculate $\alpha = exp(-(H'-H))$. H' is the recent Hamiltonian.
\item accept or reject the final $q'$ as per the acceptance probability. 
\item repeat the above steps.
\end{enumerate}

$\ell$ and $\epsilon$ are parameters of the algorithm to be tuned for best performance.
The use of HMC leads to faster convergence of MCMC sequence 
to the target pdf, as shown in the results subsection.

Some of the properties of Hamiltonian and its dynamics, like  time reversibility, volume preservation and 
time invariance (energy conservation) are important
for the application of Hamiltonian dynamics for MCMC \cite{Neal}. The reversibility is important for showing that MCMC updates that use the dynamics leave the desired distribution invariant. The acceptance probability 
of proposals found by Hamilton dynamics is one, when the Hamiltonian is conserved. 

Due to its slower convergence to the stationary density, plain MCMC is recommended only 
for special problems, where usual sampling 
methods cannot be applied readily. Here we apply HMC, which has better convergence 
properties, to simple one dimensional problems to illustrate the new approach of post adaptive
sampling method. This method requires MCMC to sample from the
truncated distributions appearing in reliability problems,
whose normalisation constant is not known. The
following section discusses the method to construct a normalised 
$pdf$ from MCMC or HMC samples. 

\subsection{Non-parametric kernel density estimators}

The procedure outlined in section \ref{sec:psam} hinges on the ability
to construct a normalised function which closely matches the sampled function. 
One of the simplest means of constructing a pre-normalised function $h(x)$, which  
has the  same support as the p.d.f we are interested in, is 
to obtain it using Kernel Density Estimator (KDE). 

Given that the samples are generated from the desired distribution,
kernel density estimator~\cite{Hansen} is used to construct the  density $h(x)$ as,
\begin{equation}
h(x,w) = \frac{1}{N w}\sum_j {\phi((x-x_j)/w)}
\label{eq:kde}
\end{equation}
where $\phi$ being the standard normal kernel,
\be
\phi(x)=\frac{1}{\sqrt{2 \pi}}e^{\frac{-x^2}{2}}
\ee
$x_j$ are the sampled data points and $w$ a bandwidth parameter controlling the
degree of smoothing. Smaller bandwidth parameter results in estimated 
density closely following given data and larger bandwidth more smooth estimate. 
Optimum $w$ is estimated using Silverman's formula ~\cite{Hansen}. 
$$
w=0.9 \,\sigma N^{-0.2}
$$
Many methods have been suggested in the past for obtaining bandwidth estimates. The 
procedures may estimate a fixed $w$ or a variable $w$~\cite{Hansen}. When estimating
variable $w$, it may be associated with $x$ or $x_j$ of equation \ref{eq:kde}. Further
generalisation would make $w$ a matrix, increasing the complexity of KDE.
   
Since $h(x,w)$ must be proportional to $g\prime(x)$ for obtaining 
minimum variance, one can define a functional,
\be
L= \sum_i \left [ h(x_i,w) - \lambda. g\prime(x_i) \right ]^2
\ee
for minimization with respect to the parameters $\lambda$ and $w$.
In this work, $w$ is kept fixed. Solving for $\lambda$ we get,
\be
\lambda =\frac{\sum_i h_i g_i}{\sum_i g_i^2}
\label{eq:21}
\ee
$\lambda$ is identified from equation (\ref{eq:10}) as the inverse of the probability estimate. 
In the above expression $g \prime$ is just denoted as $g$ as it is clear
from the context that only $g \prime$ is available for calculations.
The expression for failure probability as given by the variational 
formulation (equation \ref{eq:21}) can be compared
with the expression derived directly,  equation \ref{eq:20}.
  
The plain KDE as in eq. ~\ref{eq:kde} needs edge correction, especially if the
the density being estimated is highly asymmetric near boundaries.
Reflection and rescaling are two widely used methods~\cite{Jones} for edge 
correction. Both variants are tested for failure probability estimates
i.e., for estimation of area under left truncated $pdf$s.
In the next section we discuss the MCMC algorithm to be used for the
generation of samples from $g'(x)$. 
 
\section{Numerical experiments}

The MCMC algorithm is very general for sampling applications with the 
added advantage  that one need not know the normalising constant of the $pdf$ to generate the 
samples. However, there could be bias in the samples, due to initial part 
of the chain and may converge slowly to the target distribution.
The observed bias in generating samples from uniform and
normal distributions for about $10^4$ samples is shown in the
following figures \ref{100} and \ref{101} respectively. The bias near the peak and right edge 
is shown for truncated Normal distribution in 
figure \ref{102}, in linear scale and in figure \ref{103} in log scale.  

\begin{figure}[Htb]
\centering
\subfloat[Uniform distribution with 10K samples]{
\includegraphics[scale=0.5,width=2.0in,height=1.5in]{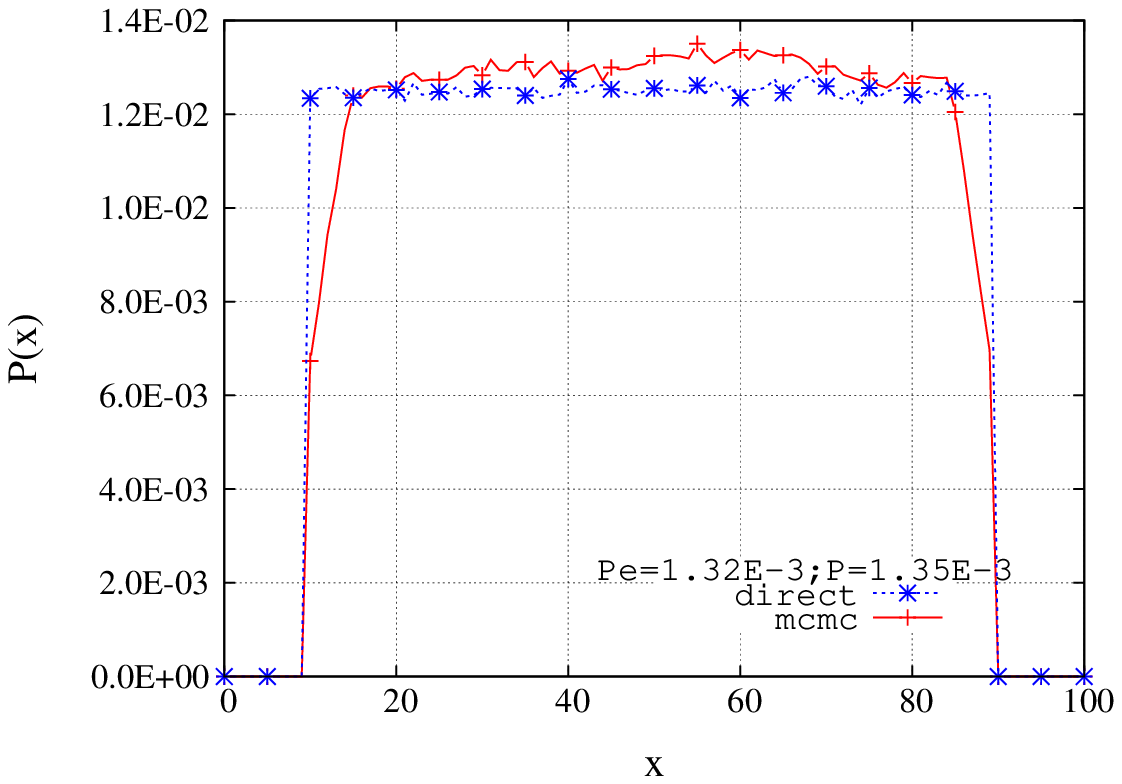}
\label{100}}
\quad
\centering
\subfloat[Gaussian distribution with 10K samples]{
\includegraphics[scale=0.5,width=2.0in,height=1.5in]{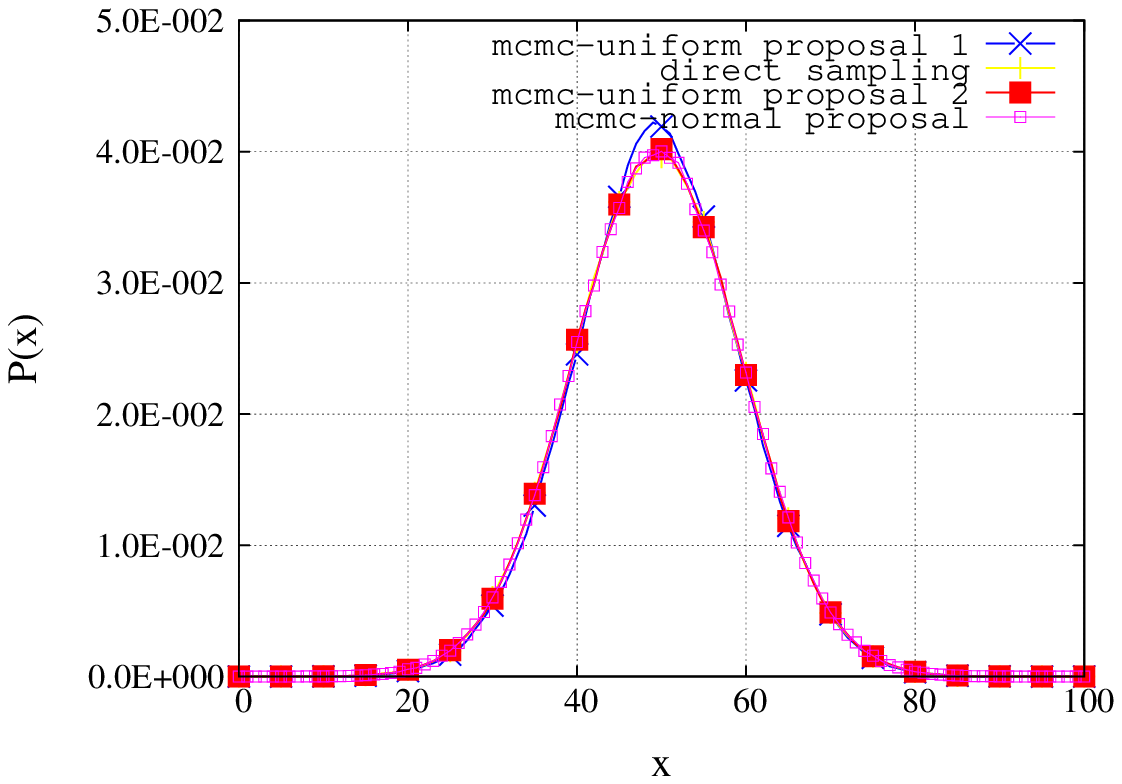}
\label{101}}

\centering
\subfloat[Truncated Normal density, in linear scale]{
\includegraphics[scale=0.5,width=2.0in,height=1.5in]{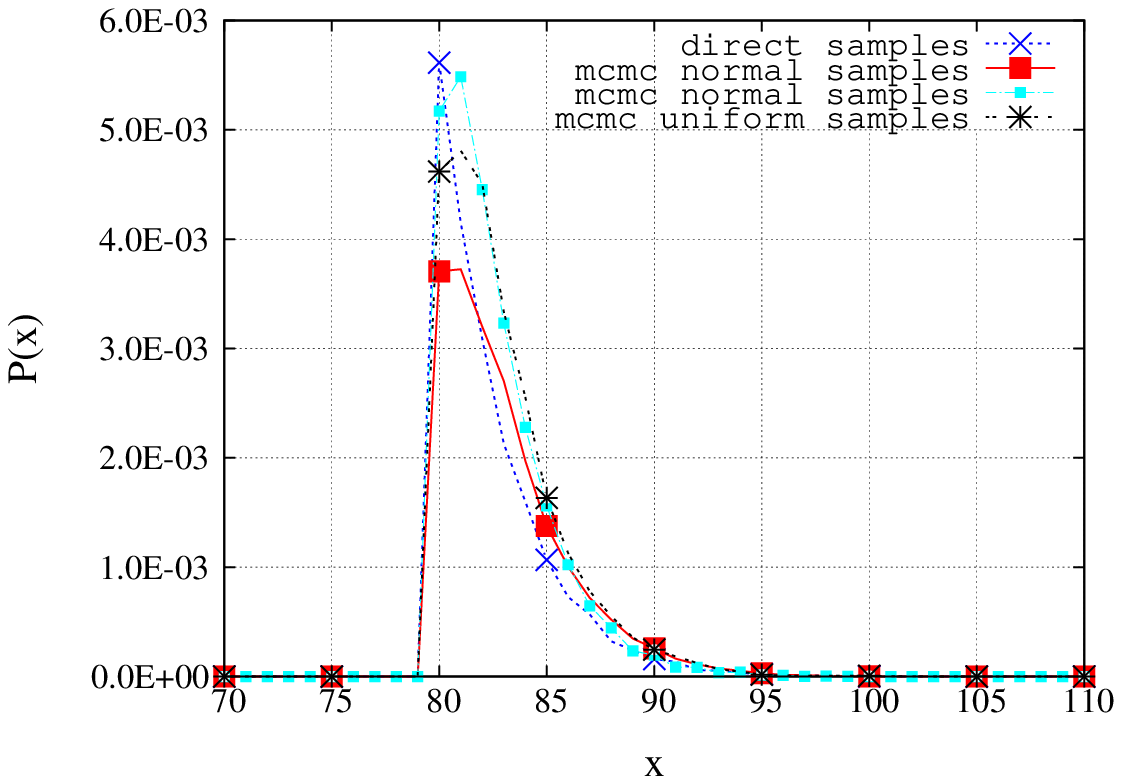}
\label{102}}
\quad
\centering
\subfloat[Truncated Normal density, in log scale]{
\includegraphics[scale=0.5,width=2.0in,height=1.5in]{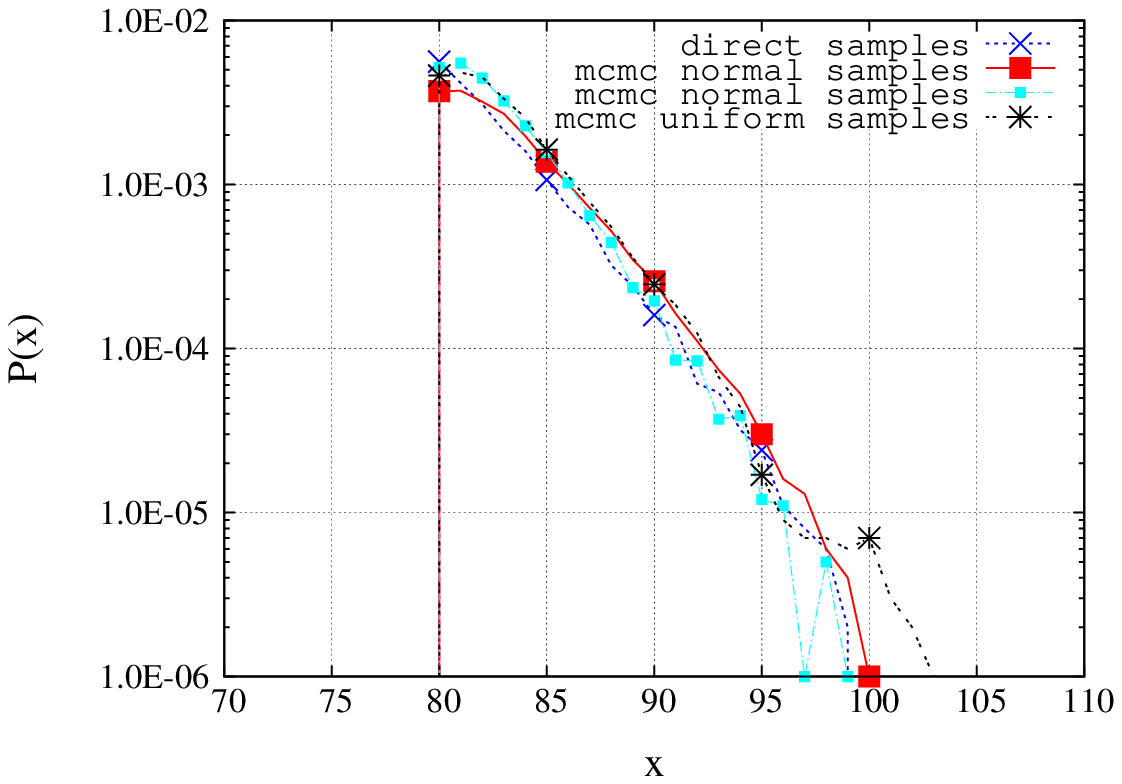}
\label{103}}
\caption{Slow convergence of plain MCMC samples near peak and edges of target densities}
\end{figure}

The application of the post sampling optimization method to the integration of a known 
function is presented here. A Gaussian function N(0,5), mean 0, standard deviation 5, 
is used to check the performance of the algorithm. 
The table~\ref{tab:psa1} presents the results for the case of variable exceeding $3 \sigma$,  and 
table~\ref{tab:psa2}  presents results for the case of variable exceeding $4 \sigma$ level.  
In the tables, results from additional set of runs are given in the last three columns.

\begin{table}[htbp]
\centering
\caption{\label{tab:psa1} Exceedence Probability with Post Sampling Variational Monte Carlo with HMC ($P \{x \geq \mu+3 \sigma\}$). }
\bigskip
\begin{tabular}{|l|p{2cm}|c|c|p{1cm}|c|c|p{1cm}|} \hline
Method & Sample size (N) & P & $\delta$ & cov ($\Delta$) & P & $\delta$ & cov ($\Delta$)  \\ \hline
Analytical & -  & 1.349E-3 & -     & -    & - & - & -  \\ \hline
MCS  & $2.10^4 \times 10$ & 1.285E-3 & -    & 13.3 & 1.390E-3 & 0.078 &  33.2 \\ \hline
MCMC   & $2.10^4 \times 10$  & 1.336E-3 & - & 28.2 & 1.337E-3 & 0.075 &  31.0 \\ \hline
IS-MCMC  & $2.10^4 \times 10$ & 1.43E-3 & -  & 11.7 & 1.367E-3 & 0.017 &  7.3  \\ \hline  
PSV-HMC & $10^4 \times 10$ &  1.337E-3 & -  & 0.04 & 1.330E-3 & 5.5E-4 & 0.05  \\ \hline
\end{tabular}
\end{table}

\begin{table}[htbp]
\centering
\caption{\label{tab:psa2} Exceedence probability with post sampling variational Monte Carlo using HMC ($P\{x \geq \mu+4 \sigma\} $) }
\bigskip
\begin{tabular}{|l|p{2cm}|c|c|p{1cm}|c|c|p{1cm}|} \hline
Method & Sample size (N) & P & $\delta$ & cov ($\Delta$) & P & $\delta$ & cov ($\Delta$)  \\ \hline
Analytical & -  &  3.1671E-5 & -  & -    & - & - & -  \\ \hline
MCS  & $2.10^4  \times 10$ & 1.0E-5 & 0.35  & 149.1 & 3.0E-5 &0.35 &  149.1 \\ \hline
MCMC   & $2.10^4  \times 10$  & 2.646E-5 &0.097  & 41.1 & 2.83E-5 &0.09 &  38.7 \\ \hline
IS-MCMC  & $2.10^4  \times 10$ & 3.209E-5 & 0.016 &6.62 & 3.131E-5 & 0.017 &  7.3  \\ \hline  
PSV-HMC & $10^4 \times 10$ & 3.12E-5   & 4.9E-4  & 0.05 & 3.165E-5 & 5.1E-4 & 0.05  \\ \hline
\end{tabular}
\end{table}

The results from Direct Monte Carlo Simulation (MCS), Simple 
Importance Sampling about the limit point using MCMC (IS-MCMC) and Post Sampling Adaptive (PSV-HMC) Hybrid Monte Carlo methods
are compared in the tables. The IS-MCMC uses 
a simple  weighting pdf, which is a Gaussian pdf with mean shifted to $x_T$ , the limit point.
The $\Delta (P)$ shown in the table is known as unitary coefficient of variation and is a 
measure of performance of the algorithm. It is independent of
number of samples $N$ but is a function of $P$. It is related to 
fractional error $\delta$ (also known as  $cov$) as $\Delta = \delta \sqrt{N} $.  

The statistical performance of the algorithm is better than simple IS Monte Carlo.
It appears from the table {\ref{tab:psa1}}, that both the importance sampling 
results are biased and is likely the result of the use of Markov Chain random number generator with
the chains not having reached equilibrium. It is to be observed that the measure $\Delta$ does not 
consider the additional computation time required for constructing 
$h(x)$ by kernel density estimation. This makes the 
proposed method slower compared to direct MCS for simpler computations of $C(x)$.

The following points are to be considered while implementing the
MCMC algorithm. It has been recommended in the
literature that optimal acceptance ratio is about $25 \%$ for MCMC algorithm.
For HMC the optimal value is about $67 \%$. 
The size of the proposal is of the order of target distribution for optimum
performance and same shape as target distribution is preferred, unless the 
proposal distribution itself is part of the optimization. 

\begin{figure}[htb]
\centering
\begin{minipage}[b]{0.4\linewidth}
\includegraphics[scale=0.45,width=2.5in,height=1.5in]{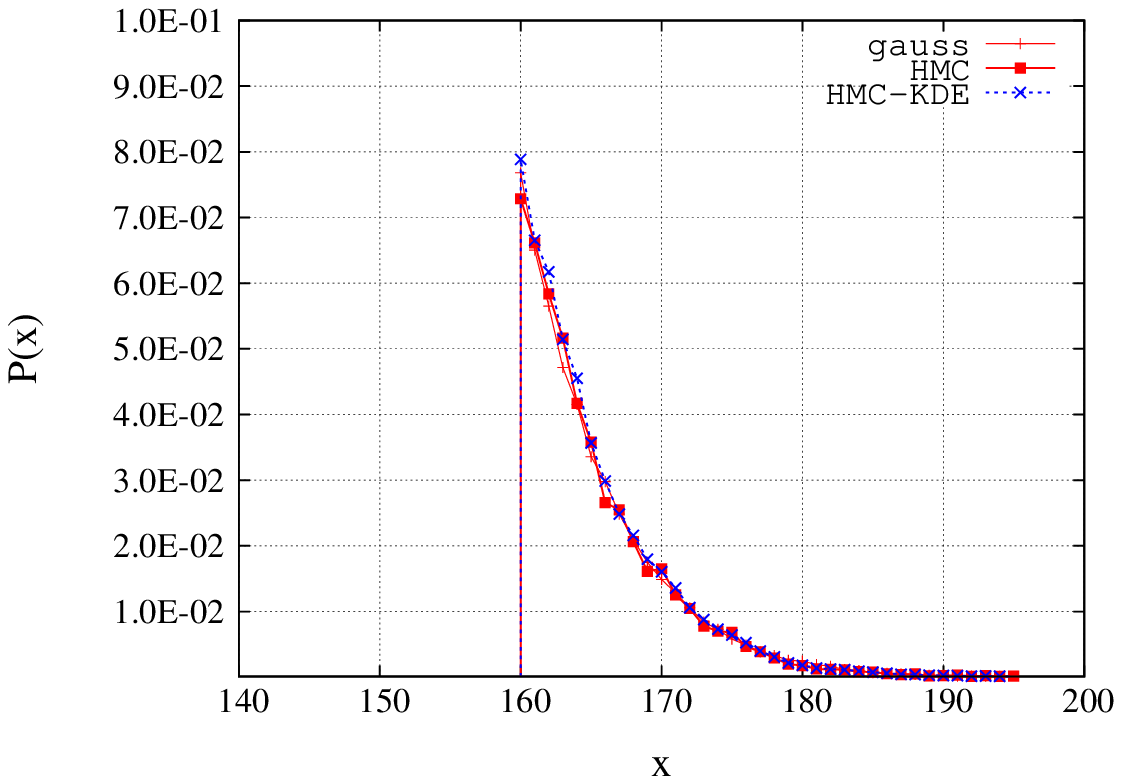}
\caption{Convergence near peak region for truncated Normal distribution
beyond 3$\sigma$, generated using HMC with edge correction.}
\label{105}
\end{minipage}
\quad
\begin{minipage}[b]{0.4\linewidth}
\includegraphics[scale=0.45,width=2.5in,height=1.5in]{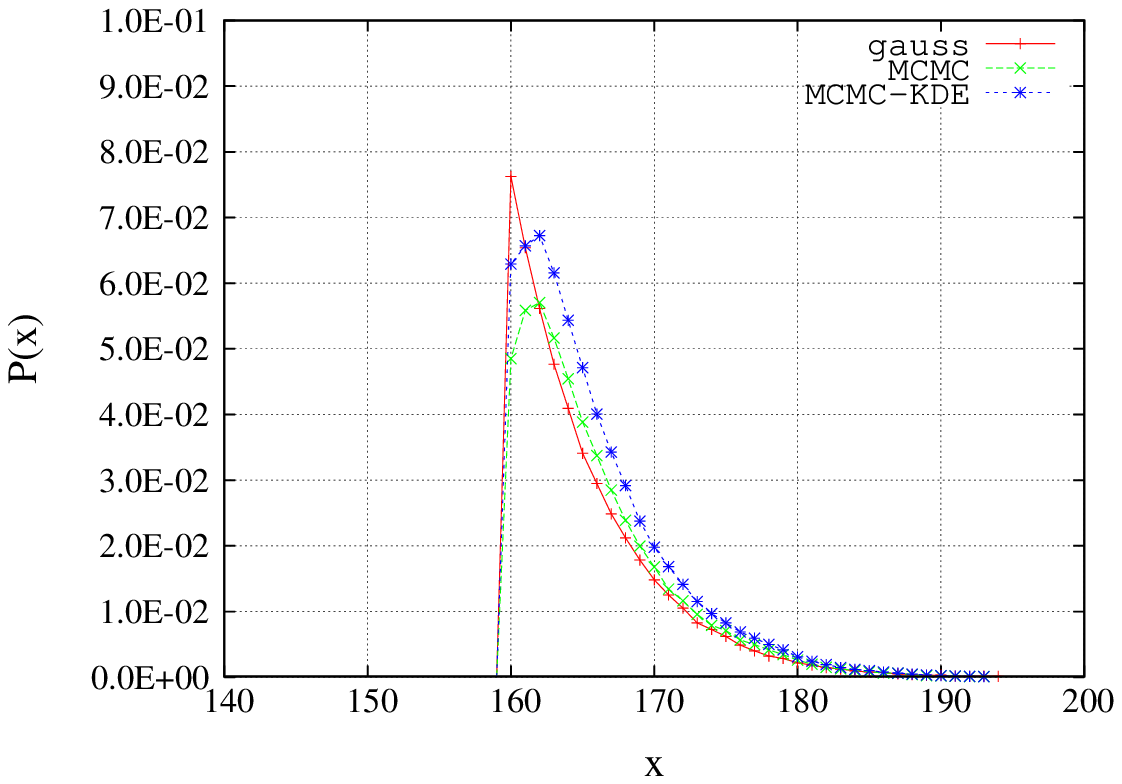}
\caption{Slow convergence near peak region for truncated Normal distribution
beyond 3$\sigma$, generated using MCMC, with edge correction.}
\label{106}
\end{minipage}
\end{figure}

\begin{figure}[htb]
\centering
\includegraphics[scale=0.9,width=5.5in,height=3.5in]{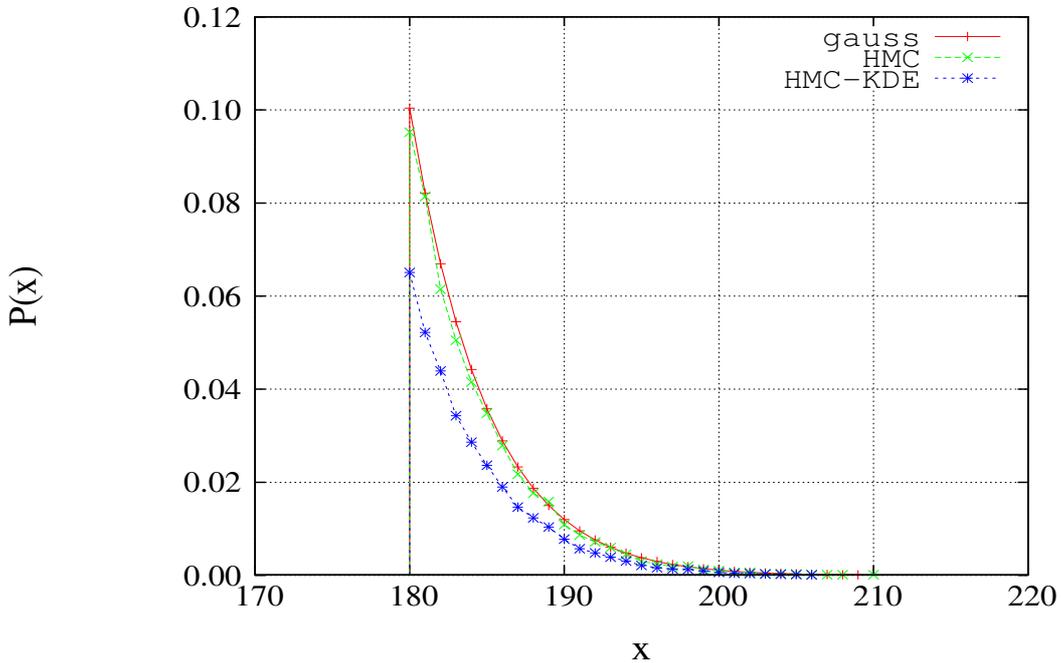}
\caption{Convergence near peak and tail regions with HMC for truncated Normal distribution
beyond 4$\sigma$ generated using HMC and edge correction. }
\label{107}
\end{figure}

The MCMC algorithm, though is convenient for generating random number
from complex distributions, is prone to bias. The relative error and coefficient of 
variation metrics might present very optimistic picture statistically. While there will be still
significant systematic error in the estimate.To alleviate the problem of slow
convergence and bias, HMC algorithm is implemented in the study. The performance of the 
HMC for truncated Normal distribution is 
shown in figure \ref{105}. The performance without HMC is  shown in figure
\ref{106} for comparison. Figure \ref{107} presents the results for the case of
exceedence limit set at 4 $\sigma$ for HMC case.

In figures, \ref{105} and \ref{106} 'gauss' or 'standard samples' mean  that the samples have 
been derived using \textit{python's} (programming language)  Gaussian random number generator 'gauss" or analytical functions. \textit{Python} uses Mersenne Twister as the core random number generator, producing 53 bit precision floats with period $2^{19937}-1$. 

\begin{figure}[htb]
\centering
\subfloat[Adaptive
MCMC algorithm, with $10^3$ samples repeated 10 times]{
\includegraphics[width=2.5in,height=1.8in]{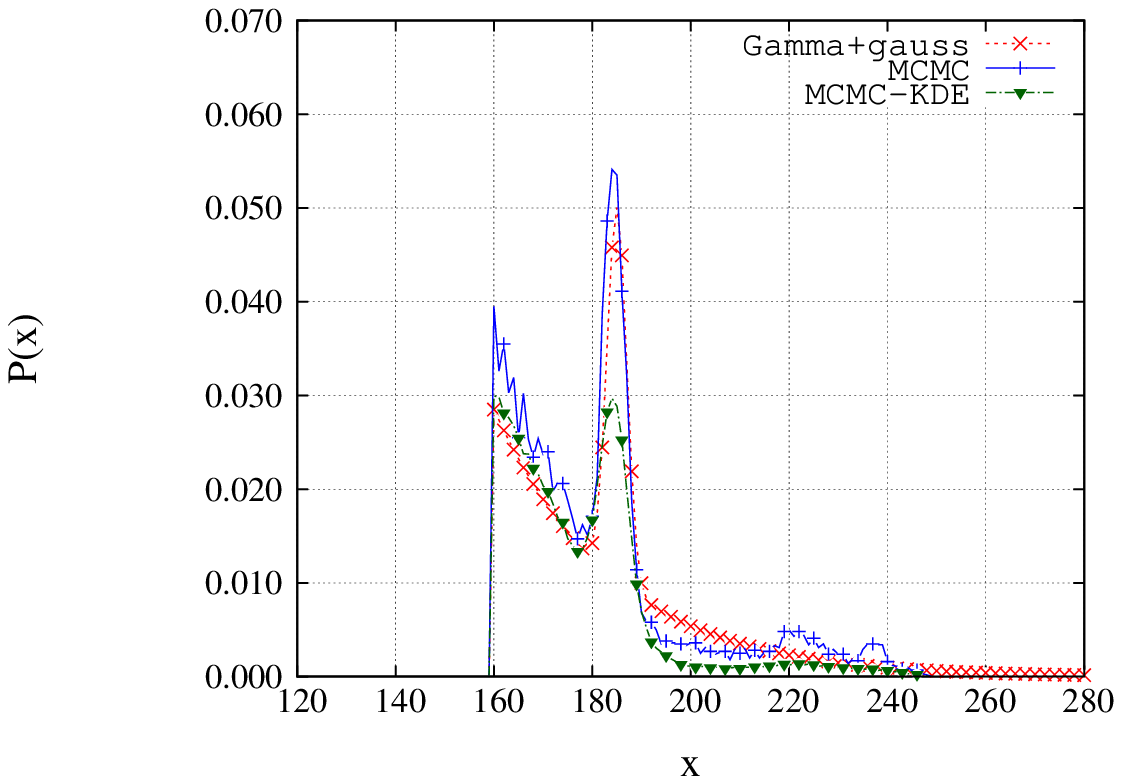}
\label{108} 
}
\quad
\subfloat[Adaptive
HMC algorithm, with $10^3$ samples repeated 10 times]{
\includegraphics[width=2.5in,height=1.8in]{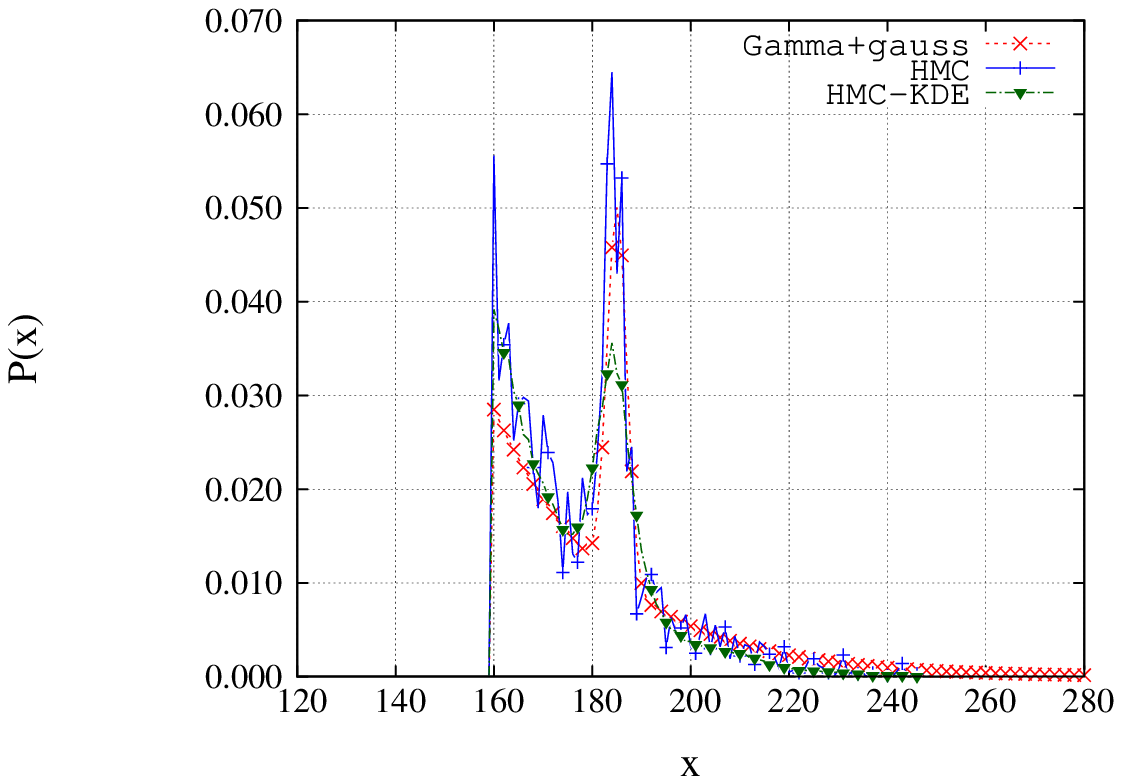} 
\label{109}
}
\caption{Convergence near the truncated Gamma and Gaussian peaks using MCMC and HMC algorithms}
\label{1080} 
\end{figure}

Next the convergence of the adaptive MCMC and  adaptive HMC 
were tested for a mixture of truncated Gamma and Gaussian distribution. The mixed distribution is obtained 
as,
$$
g(x;\alpha,\nu) =f_1 \frac{\alpha^n}{\Gamma(n)}e^{-\alpha x} x^{\nu-1} + f_2  N(\mu,\sigma)
$$
with $\alpha = 0.05$, $\nu=2.5$, $\mu=185$, $\sigma=2$, $x_T=130$, and  $f_1=0.998$. $ N(\mu,\sigma)$ denotes normal distribution with mean $\mu$ and standard deviation $\sigma$. 
The convergence with adaptive  MCMC is depicted in figure \ref{108} and the convergence
with adaptive HMC is shown in figure \ref{109}. Though visually, KDE in the two figures look
more or less same, quantitative results indicate better performance by HMC as seen
from table \ref{tab:psa3}.  In this case of mixed distribution with two peaks in the domain of integration,
MCS out performs ISMCS. This is due to the tail of the distribution being less skewed than the normal tail. 
 
\begin{table}[htbp]
\centering
\caption{\label{tab:psa3} Exceedence probability for mixed distribution with post sampling variational Monte Carlo using HMC ($P \{x \geq x_T\} $) }
\bigskip
\begin{tabular}{|l|p{2cm}|c|c|p{1cm}|c|c|p{1cm}|} \hline
Method & Sample size (N) & P & $\delta$ & cov ($\Delta$) & P & $\delta$ & cov ($\Delta$)  \\ \hline
Analytical & -  &  8.8294E-3 & -     & -    & - & - & -  \\ \hline
MCS  & $10^4\times 10$ & 8.845E-3 & 0.012  & 3.63 & 8.871E-3 & 8.1E-4 &  2.5 \\ \hline
MCMC   & $10^4\times 10$  & 11.5E-3 & 0.132 & 39.7 &10.857E-3 & 0.12 &  35.7 \\ \hline
IS-MCMC  & $10^4\times 10$ & 9.475E-3 & 0.057  &16.9 & 8.277E-3 & 0.053 &  9.5  \\ \hline  
PSV-HMC* & $5.10^3\times 10$ &  9.33E-3 & 1.4E-3  & 0.1 & 9.179E-3 & 1.2E-3 & 0.1  \\ \hline
\end{tabular}
\flushleft
*The average of the two estimates by the formulae (equations \ref{eq:20} and  \ref{eq:21}) for a 
few cases of $10^4$ runs are,
8.64E-3, 8.83E-3, 8.612E-3,7.53E-3,  8.7E-3. While one estimate is biased lower and another estimate biased higher, the average seem to be better for this distribution.
\end{table}

The efficiency of the algorithm has been compared only with respect to the number of 
samples required by different methods. The method of using KDE entails $O(N*m)$ worst case 
sampling operations in one dimension, where $N$ is the number of samples in the MCMC
step and $m$ is the 
number of samples falling within the range of the kernel density. Therefore
speed up in running time will be fully realized only when the computational cost of $C(x)$
evaluation is  significantly higher than the evaluation of probability densities. 
 
\clearpage
\section{Conclusion}
A simple, and flexible algorithm for  importance sampling 
has been presented. In this approach, the optimization 
problem in the evaluation of the integral is shifted to the function construction
phase than at the sampling phase, which is the case in the current methods. The 
proposed importance sampling is quite general as there is no manual selection
of matching densities in the sampling phase. The method has been tested with Gaussian
and mixture of Gamma and Gaussian densities
for convergence and error. The method is observed to perform better compared to 
well known importance sampling methods. The method's performance can be further
improved by better variational optimization strategies.
\bigskip

\textbf{Acknowledgements:}
The author thanks director RDG, IGCAR, Dr. P. Chellapandi for his encouragement and support during the course of this work.
\begin{singlespace}
\bibliographystyle{unsrt}
\bibliography{mcs-ref}
\end{singlespace}

\end{document}